\documentclass[11pt]{article}
\usepackage{moriond,epsfig}

\def\deg{^{\circ}}

\begin{document}
\vspace*{4cm}
\title{NEW RESULTS FROM CHANDRA}

\author{ W. FORMAN, M. MARKEVITCH, \& C. JONES}

\address{Smithsonian Astrophysical Observatory\\
60 Garden St. Cambridge, MA 02138}

\author{ A. VIKHLININ}

\address{Smithsonian Astrophysical Observatory\\
60 Garden St. Cambridge, MA 02138\\
\& Space Research Institute\\
Profsouznaya 84/32\\ 
Moscow 117810, Russia}

\author{E. CHURAZOV}
\address{MPI fur Astrophysik\\
Karl-Schwarzschild-Strasse 1\\
Garching, Germany\\
\& Space Research Institute\\
Profsouznaya 84/32\\ 
Moscow 117810, Russia}

\maketitle 

\abstracts{We discuss two themes from Chandra observations of galaxies
and clusters. First, we describe the effects of radio-emitting plasmas
or ``bubbles'', inflated by active galactic nuclei, on the hot X-ray
emitting gaseous atmospheres in galaxies and clusters. We describe the
interaction of the ``bubbles'' and the X-ray emitting gas as the
buoyant bubbles rise through the hot gas.  Second, we describe sharp,
edge-like surface brightness structures in clusters. Chandra
observations show that these features are not shock fronts as was
originally thought, but ``cold fronts'', most likely the boundaries of
the remaining cores of merger components. Finally, we present recent
observations of M86 and NGC507 which show similar sharp features around
galaxies. For M86, the sharp edge is the boundary between the galaxy's
X-ray corona and the Virgo cluster gas. The structures around NGC507,
the central galaxy in a group, could be relics of galaxy formation or
may reflect the motion of NGC507 in the larger potential of the group.
}

\section{Introduction}

\vspace*{0.25in}

\begin{center}
\begin{minipage}{4.0in}
{\em  Like a hell-broth boil and bubble\\
Fire burn and cauldron bubble }\\
{\footnotesize Macbeth -- Act 4 Scene 1}
\end{minipage}
\end{center}

\vspace*{0.25in}

With its first images, the Einstein Observatory changed our view of
clusters and galaxies.  In clusters we saw rich substructure
reflecting complex gravitational potentials and double clusters,
systems in the process of merging on a time scale of $\sim10^9$
yrs.\cite{jones1979,jones1984,wrf1} Elliptical galaxies, rather than
being gas poor, were found to have significant and extensive hot
atmospheres.\cite{wrf2} ROSAT and ASCA continued the revolution with
studies of cluster merging and substructure including Coma, A2256,
A754, Cygnus A, Centaurus, A1367,and Virgo, to mention just a
few.\cite{briel1991,briel1992,bohringer1994,alexey2,henry1995,honda1996,henriksen1996,alexey1,chur1999,mark1999,donnelly,schindler1999}

Chandra's high angular resolution has brought us a new view of old
friends -- early type galaxies and clusters of galaxies -- that we
thought we knew pretty well. We are familiar with the ingredients,
galaxies, radio emitting plasma, hot gas, and dark matter. The recipe
is simple -- {\em mix vigorously}. With these simple instructions we
find new and unexpected phenomena in the Chandra observations.

\section{The Radio---X-ray Connection -- or Bubbles, Bubbles Everywhere}

Prior to the launch of Chandra, ROSAT observations of NGC1275 and M87
provided hints of complex interactions between radio emitting plasmas
ejected from AGN within the nuclei of dominant, central cluster
galaxies.\cite{boh1993,boh1995,chur2000a} With the launch of Chandra,
the interaction between the radio emitting plasma and the hot
intracluster medium (ICM) has been observed in many systems and now can
be studied in detail.

\subsection{Hot Plasma Bubbles in Cluster Atmospheres}

One of the first, and clearest, examples of the effect of plasma
bubbles on the hot intracluster medium was found in the Perseus
cluster around the bright active, central galaxy NGC1275 (3C84). First
studied in ROSAT images,\cite{boh1993} the radio emitting cavities to
the north and south of NGC1275 are clearly seen in the Chandra images
with bright X-ray emitting rims surrounding the cavities that
coincide with the inner radio lobes.\cite{fabian2000} For
NGC1275/Perseus, the radio lobes are in approximate pressure
equilibrium with the ambient, denser and cooler gas and the bright
X-ray rims surrounding the cavities are softer than the ambient gas.
Therefore, the radio cavities are not likely to be a major source of
shock heating and hence, energy input.  The central galaxy in the
Hydra A cluster also harbors X-ray cavities associated with radio
lobes that also show no evidence for shock
heating.\cite{mcnamara2000} Both sets of radio bubbles, being of lower
density than the ambient gas, must be buoyant.

\begin{figure} [bth]
\centerline{\includegraphics[width=0.40\textwidth]{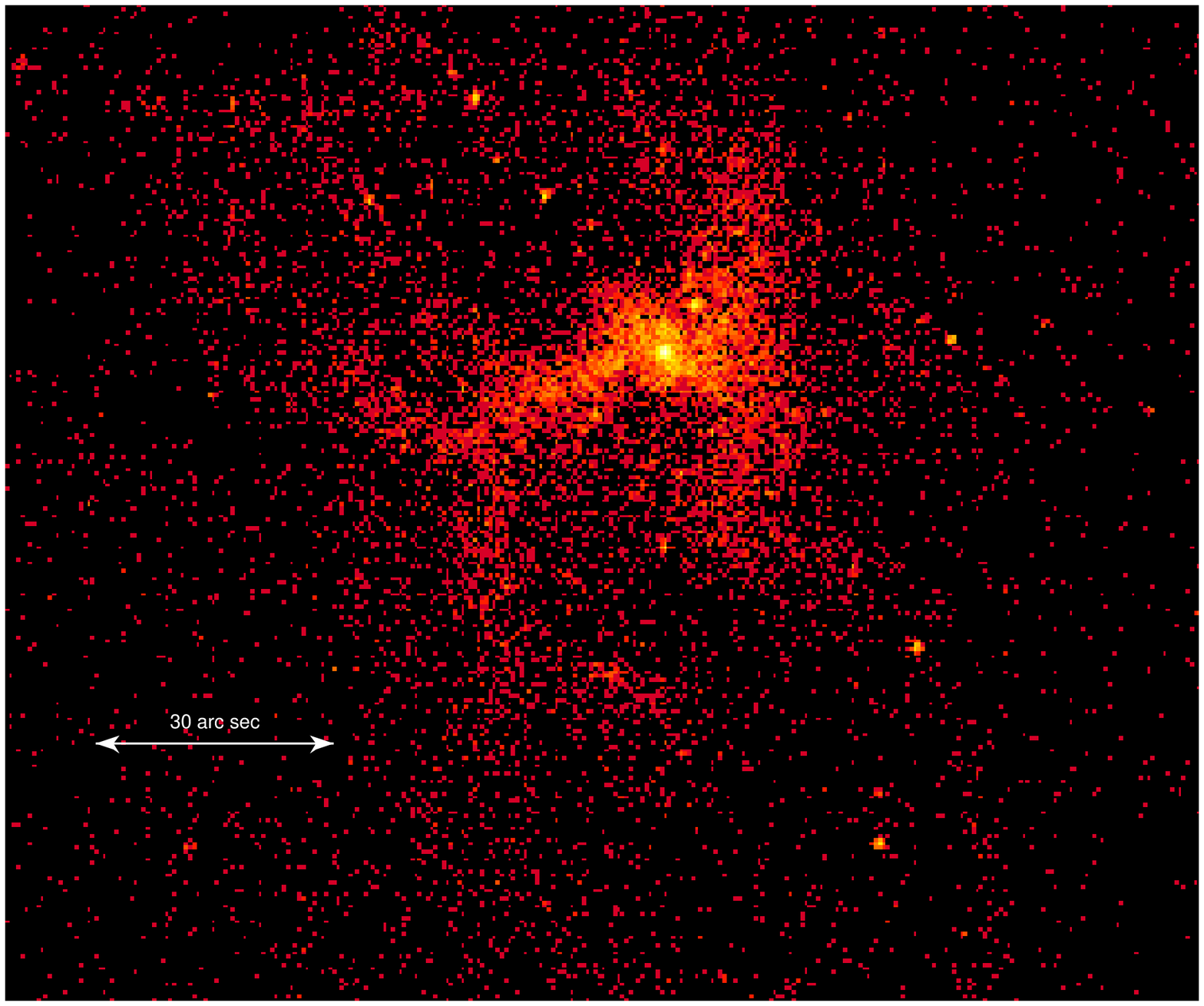}
\includegraphics[width=0.50\textwidth]{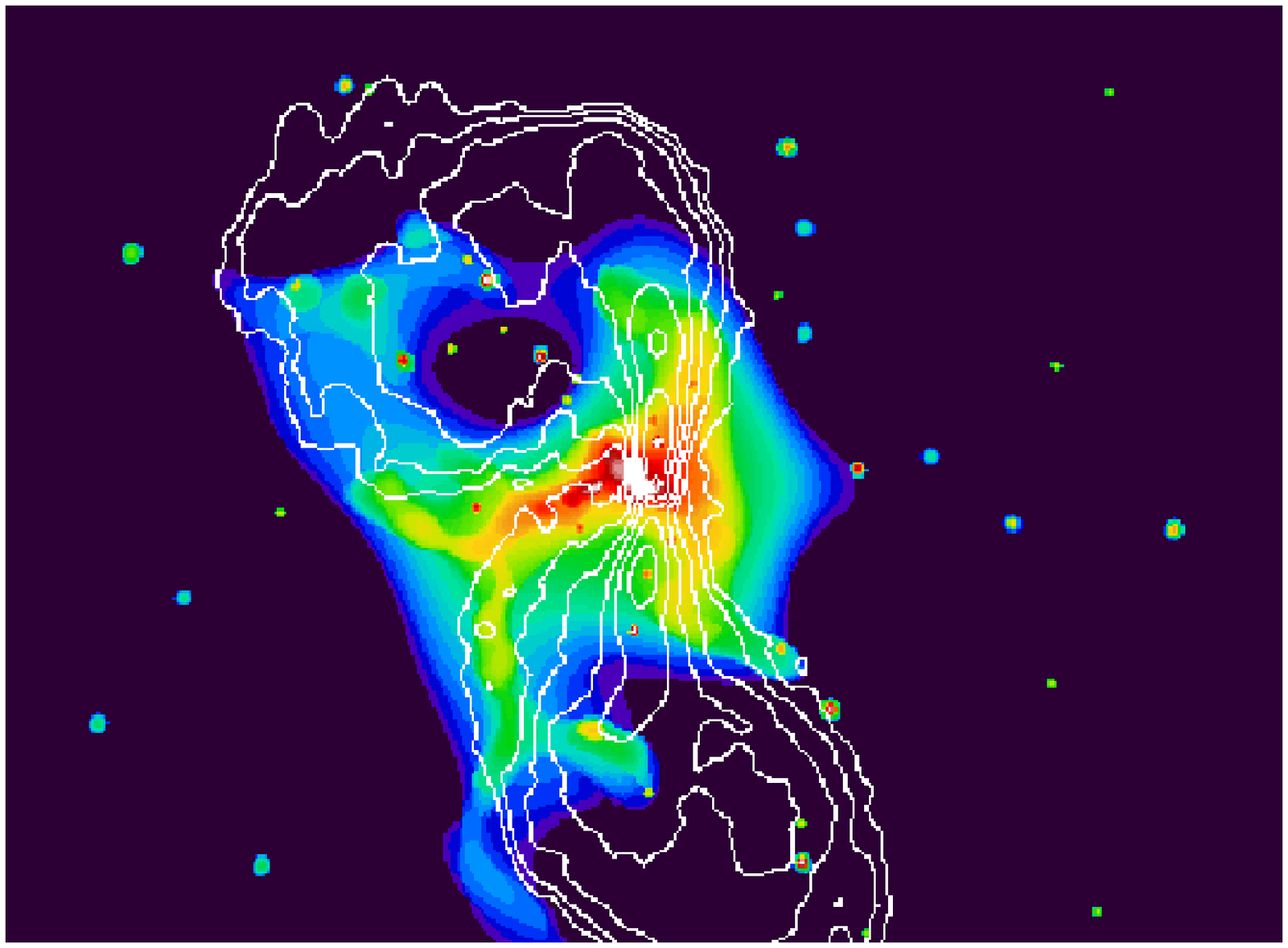}}
\caption{Left: ${\cal H}$-shaped X-ray emission from M84 image
(0.5--2.0 keV band) as observed with Chandra. The X-ray emission shows a
bar extending east-west and two approximately perpendicular filaments.
The galaxy nucleus lies in the bright region at the west end of the
bar. Right: Radio emission contours overlayed on the smoothed Chandra
image ``explains'' the unusual X-ray
morphology -- the radio plasma has displaced the hot X-ray emitting
gas. See Finoguenov \& Jones (2001) for details. }
\label{m84}
\end{figure}

The Chandra images of Perseus/NGC1275 also suggest the presence of
older bubbles produced by earlier outbursts.\cite{fabian2000} These
older bubbles appear as X-ray surface brightness ``holes'', but unlike
the inner bubbles, these outer holes show no detectable radio
emission, suggesting that the synchrotron emitting electrons may have
decayed away leaving a heated, plasma bubble.  Such bubbles, with no
attendant radio emission, are seen by Chandra in the galaxy groups
HCG62 and MKW3s.\cite{vrtilek2001,mazzotta2001}

\subsection{Bubbles In A Galaxy Atmosphere -- M84} 

The examples above focus on the radio--X-ray connection
around central galaxies in clusters and groups.  These galaxies are
clearly the brightest cluster members and occupy a special position in
the cluster gravitational potential. However, a notable example of the influence of
radio plasma on the X-ray emitting gas in a more typical early-type
galaxy is M84 (NGC4374), an E1 galaxy, within, but not at the center,
of the core of the Virgo cluster.

Finoguenov \& Jones found a very complex structure in the X-ray
emitting gas around the Virgo galaxy M84 (NGC4374) whose unusual X-ray
morphology is explained by the effect of the radio lobes on the
hot gas.\cite{finoguenov2001} Fig.~\ref{m84} shows the strong
influence of radio bubbles on the X-ray emitting gas distribution. The
X-ray emission appears ${\cal H}$-shaped, with a bar extending
east-west with two nearly parallel filaments perpendicular to this
bar.  The complex X-ray surface brightness distribution arises from
the presence of two radio lobes (approximately north and south of the
galaxy) that produce two low density regions surrounded by higher
density X-ray filaments. As with Perseus/NGC1275 and Hydra A, the
filaments, defining the ${\cal H}$-shaped emission, have gas
temperatures comparable to the gas in the central and outer regions of
the galaxy and hence argue against any strong shock heating of the
galaxy atmosphere by the radio plasma.  By determining the gas density
surrounding the radio lobes, Finoguenov \& Jones determined the
strength of the magnetic field using the observed Faraday
rotation. They inferred a line-of-sight magnetic field of $0.8 \mu$
Gauss.\cite{finoguenov2001}

In summary, the high resolution Chandra image of M84 shows the
remarkable interaction between the radio plasma and the X-ray emitting
interstellar medium (ISM) in a ``normal'' early type galaxy.  The
radio lobes have created cavities in the ISM that are surrounded by
higher density shells and allow a calculation of the magnetic field
overlying the radio bubbles.\cite{finoguenov2001}

\subsection{Evolution of Buoyant Plasma Bubbles in Hot Gaseous Atmospheres}

\begin{figure} [ht]
\centerline{\includegraphics[width=0.45\textwidth]{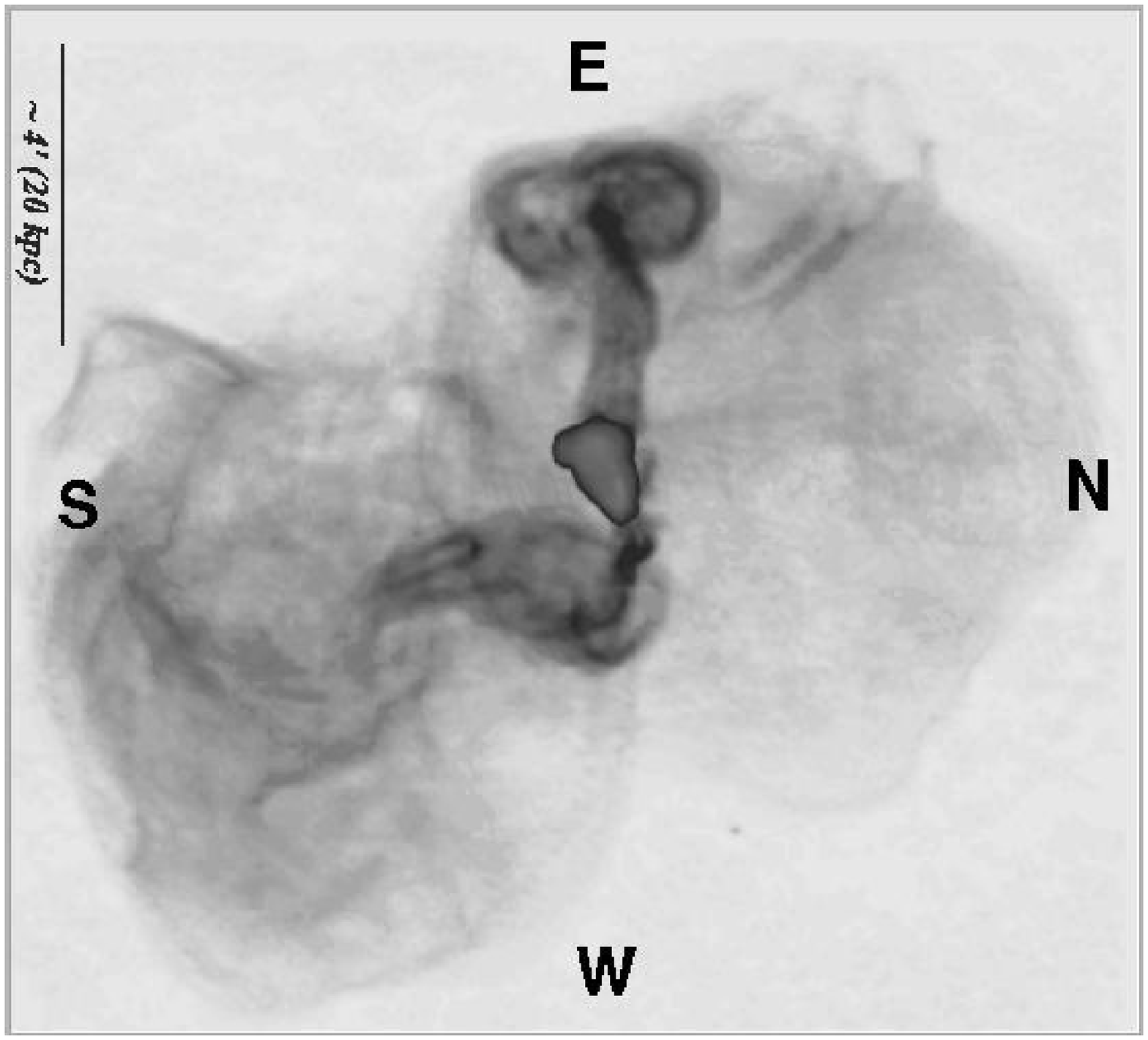}\includegraphics[width=0.45\textwidth]{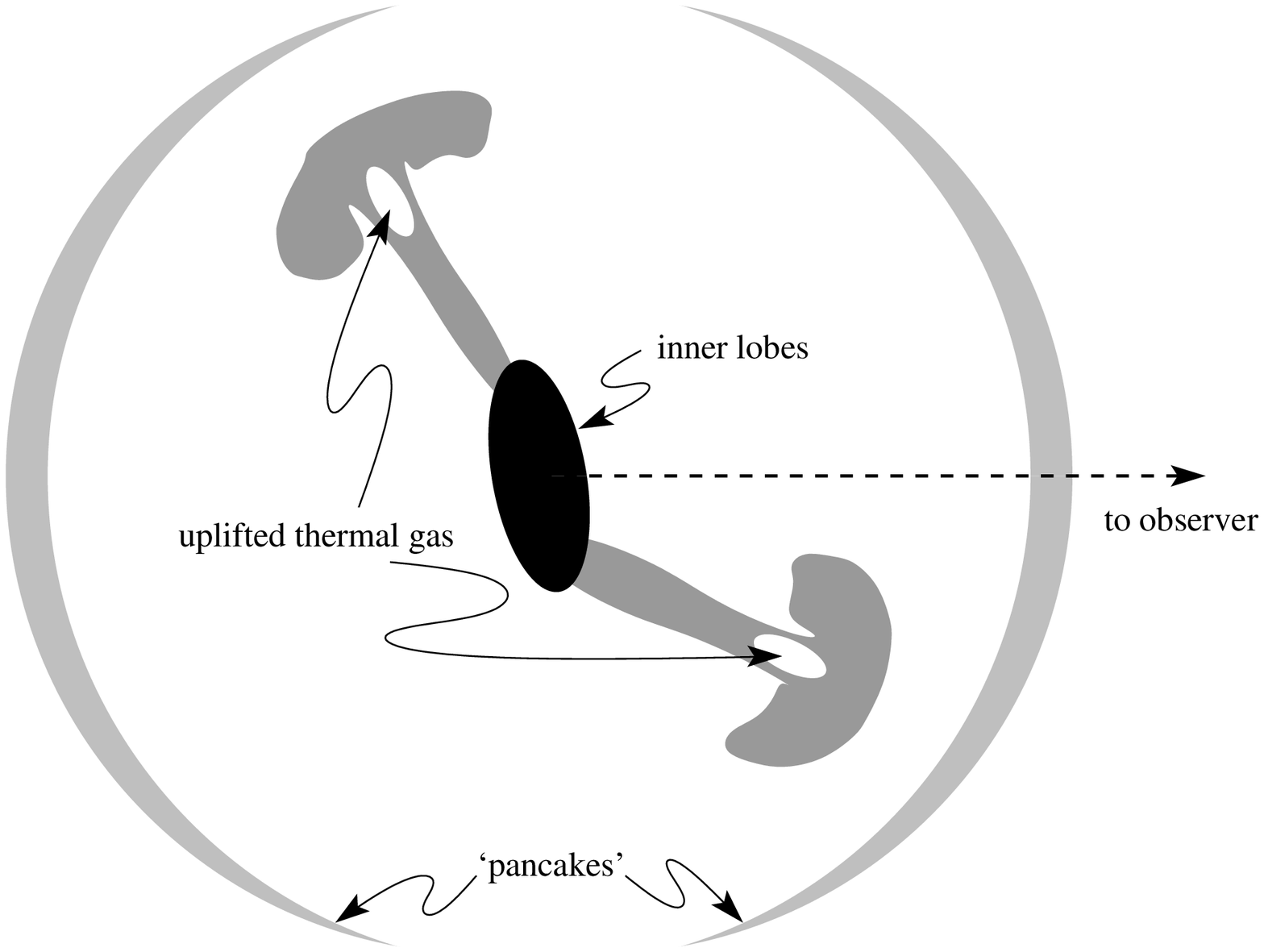}}

\caption{{\bf Left:} $14'.6 \times 16'.0$ radio map  of
M87 (North to the right, East
is up) (from Owen et al. 2000). 
{\bf Right:} Suggested source geometry.
The central black region denotes the inner radio
lobes, the gray ``mushrooms'' correspond to buoyant bubbles already
transformed into tori, and the gray lens-shaped structures are 
``pancakes'' (seen edge-on) possibly formed by older bubbles. See
Churazov et al. (2000b).
}
\label{m87}
\end{figure}

The new 327 MHz high resolution, high dynamic range radio map of M87
is shown in Fig.~\ref{m87} (left panel).\cite{owen2000}  The
high surface brightness center is the inner lobe structure (oriented
approximately north-south) with the famous jet pointing west
north-west (approximately bottom--right for the orientation of images
adopted in Fig.~\ref{m87}). Surrounding this, the highly structured
outer halo is much fainter and consists of the torus-like eastern
bubble, the much less well-defined western bubble, both of which are
connected to the central emission by a column, and the two very faint
almost circular emission regions northeast and southwest of the
center.  The correlation between X--ray and radio emitting features
has been remarked by several
authors.\cite{feigelson1987,boh1995,harris1999} The simplest
explanation for this correlation, that the excess X--ray emission is
produced by inverse Compton scattering of cosmic microwave background
photons by the same relativistic electrons that produce the
synchrotron radio emission,\cite{feigelson1987} is not supported by
the more recent observations. In particular, the ROSAT PSPC
observations show that the excess emission has a thermal spectrum and
the X--ray emitting gas in these regions has a lower temperature than
that in the surrounding regions.\cite{boh1995}

Churazov et al. proposed an alternative explanation utilizing buoyant
bubbles.\cite{chur2000b} The ``torus--like'' radio features are
strikingly similar to hot buoyant bubbles formed by powerful nuclear
atmospheric explosions.  Initially a spherical bubble is formed which
transforms into a torus and appears as a characteristic ``mushroom''
cloud as the bubble rises in the ambient medium. Another property of
powerful atmospheric explosions and buoyant bubbles is that, as the
bubble transforms to a torus, the rising bubble/torus entrains and
uplifts gas.  This may qualitatively explain the correlation of the
radio and X--ray emitting plasmas and naturally accounts for the
thermal nature of the excess emission. Finally, in the last
evolutionary phase of an atmospheric explosion, the bubble reaches a
height at which the ambient gas density equals that of the bubble.
The bubble then  expands to form a thin layer
(a ``pancake'').  The large low surface brightness features in the
radio map could be just such pancakes -- the final evolutionary phase
of the bubbles (see Fig.~\ref{m87}).  A sketch of a possible overall
source structure of M87, based on the evolution of buoyant bubbles, is
shown in Fig.~\ref{m87}.\cite{chur2000b}

Churazov et al. simulated their qualitative picture described
above.\cite{chur2000b} A spherical bubble was inflated in an
atmosphere defined by the gravitational potential of M87.  The bubble
density is 1/100 of the ambient density, making the bubble buoyant,
and the bubble temperature was 100 times the ambient value,
establishing pressure equilibrium.  The simulations showed that
buoyant bubbles behaved just as expected and did produce the features
observed in both X-rays and radio for M87. Although the exact form of
the rising bubbles was sensitive to initial conditions, the toroidal
structures were a common feature. Ambient gas was uplifted in the
cluster atmosphere reducing the effects of gas cooling and flowing to
the center and producing the ``stem'' of the mushroom that is
brighter than the surrounding regions.

Further detailed study of the X-ray and radio interaction will show
how widely the buoyant bubble scenario applies.  More detailed
simulations will provide models for more sophisticated comparisons of
theory and observation.

\section{A New Aspect of Cluster Mergers -- Cold Fronts}

For many years clusters were thought to be dynamically relaxed systems
evolving slowly after an initial, short-lived episode of violent
relaxation. However, in a prescient paper, Gunn and
Gott argued that, while the dynamical timescale for the
Coma cluster, the prototype of a relaxed cluster, was comfortably less
than the Hubble time, other less dense clusters had  dynamical
timescales comparable to or longer than the age of the Universe.\cite{gunn1972} Gunn
and Gott concluded that ``The present is the epoch of cluster
formation''.  With the launch of the Einstein Observatory came the
ability to ``image'' the gravitational potential around clusters. Many
papers in the 1980's, exploited the imaging capability of the Einstein
Observatory and showed the rich and complex structure of galaxy
clusters. 

\begin{figure}[bht]
\centerline{\includegraphics[width=0.60\textwidth]{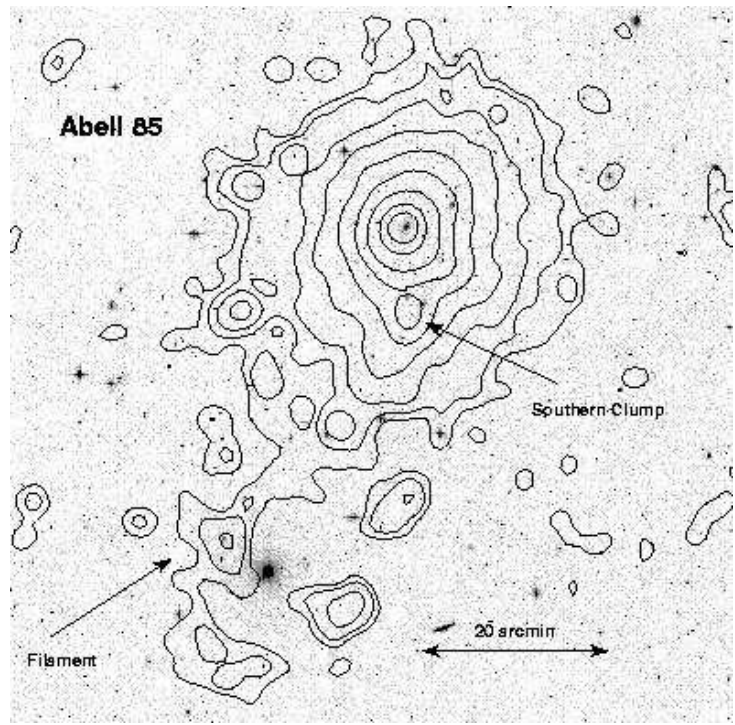}}
\caption{Optical digitized sky survey of the A85 region, with the
isophotes of the ROSAT PSPC image (0.4-2.0 keV) superposed is
shown. The X-ray contours show the
filamentary structure extending to the south southeast. The cD galaxy at the
X-ray cluster center (peak of the X-ray emission) is clearly visible
on the optical map.  The alignment of the X-ray filamentary structures
has the same position angle as the central cD galaxy, the bright
cluster member galaxies, and a much larger structure of clusters and
groups spanning $5\deg$ on the sky, almost 30 Mpc at the redshift
($z=0.0555$) of A85 (see Durret et al. 1998 for details).}
\label{a85}
\end{figure}

The X-ray observations supported the now prevalent idea that structure
in the Universe has grown through gravitational amplification of small
scale instabilities or hierarchical clustering. At one extreme, some
clusters grow, in their final phase, through mergers of nearly equal
mass components. Such mergers can be spectacular events involving
kinetic energies as large as $\sim 10^{64}$ ergs, the most energetic events
since the Big Bang. More common are smaller mergers and accretion of
material from large scale filaments.  An example showing the
relationship between large scale structure and cluster merging is
seen in the ROSAT image of A85.\cite{durret1998} Fig.\ref{a85}
shows A85 with several small groups and a merging component (small
concentration due south of the main peak). A85 exhibits common
alignments on scales from 100 kpc to 25 Mpc that are expected if
clusters form through accretion of matter from
filaments.\cite{haarlem1993} The central cluster galaxy, the bright
cluster galaxies, X-ray filamentary structure (see Fig.~\ref{a85}),
and nearby groups and clusters all show a common alignment at a
position angle of about $160\deg$.\cite{durret1998}

Chandra's high angular resolution has further illuminated the merging
process and the complexity of the X-ray emitting intracluster medium
(ICM).  Prior to the launch of Chandra, sharp gas density
discontinuities had been observed in the ROSAT images of
A3667\cite{mark1999} and in A2142 (unpublished). Since both clusters
exhibited characteristics of major mergers, these features were
expected to be shock fronts. However, the first Chandra observations
showed that these were not shocks, but a new kind of structure -- cold
fronts.\cite{mark2000} Their study has provided new and detailed
insights into the physics of the ICM.\cite{vik2001a,vik2001b}.

\subsection{Multiple Cold Fronts in A2142}

A2142 is a hot ($kT \sim 9$ keV), X-ray-luminous cluster at a redshift
of $z=0.089$. Two bright elliptical galaxies, whose velocities differ
by $1840$ km s$^{-1}$ lie near the center and are aligned in the
general direction of the X-ray brightness elongation. All these
properties suggest an unrelaxed cluster.\cite{mark2000}

\begin{figure}
\centerline{\includegraphics[width=0.50\textwidth]{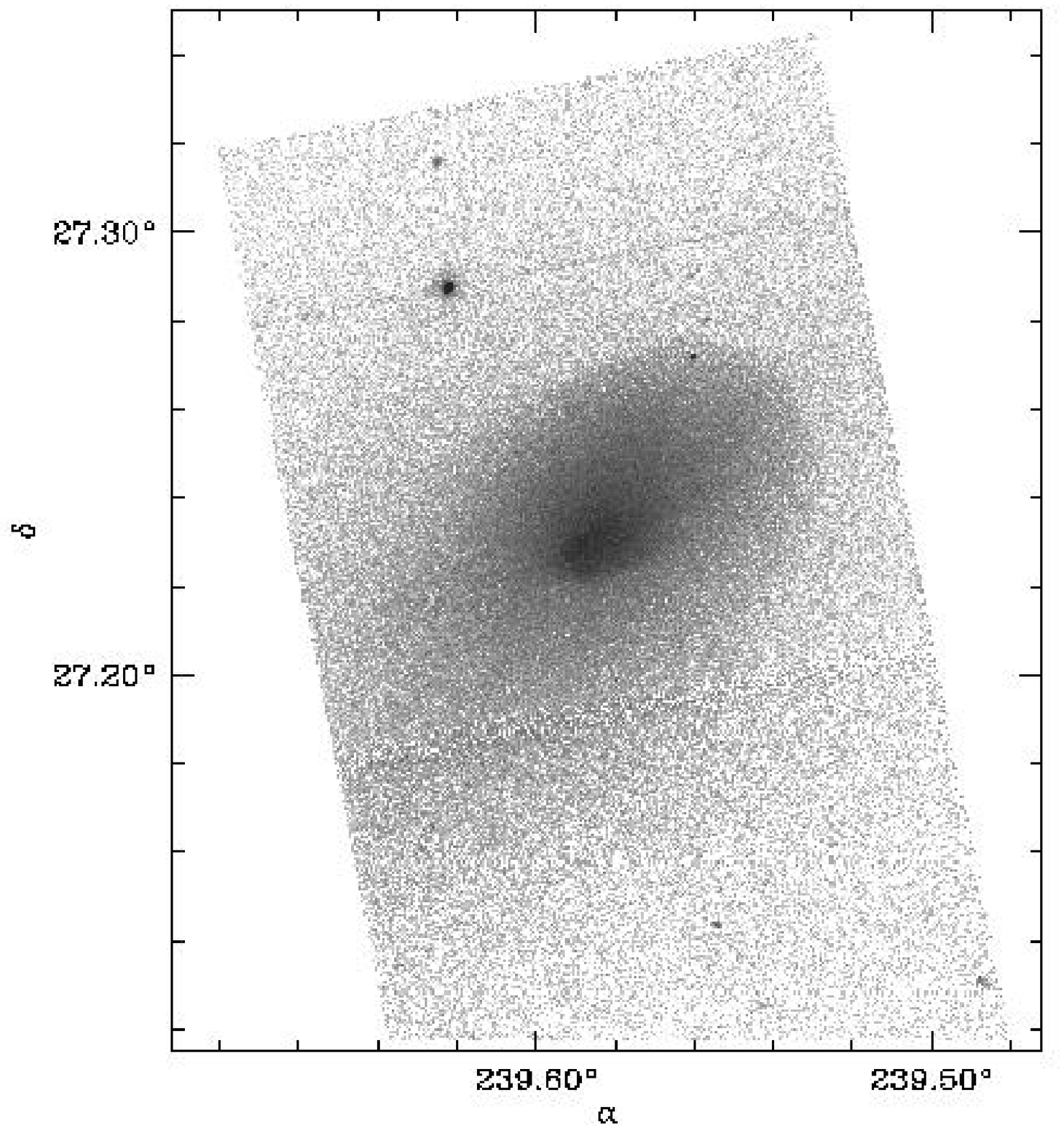}
\includegraphics[width=0.50\textwidth]{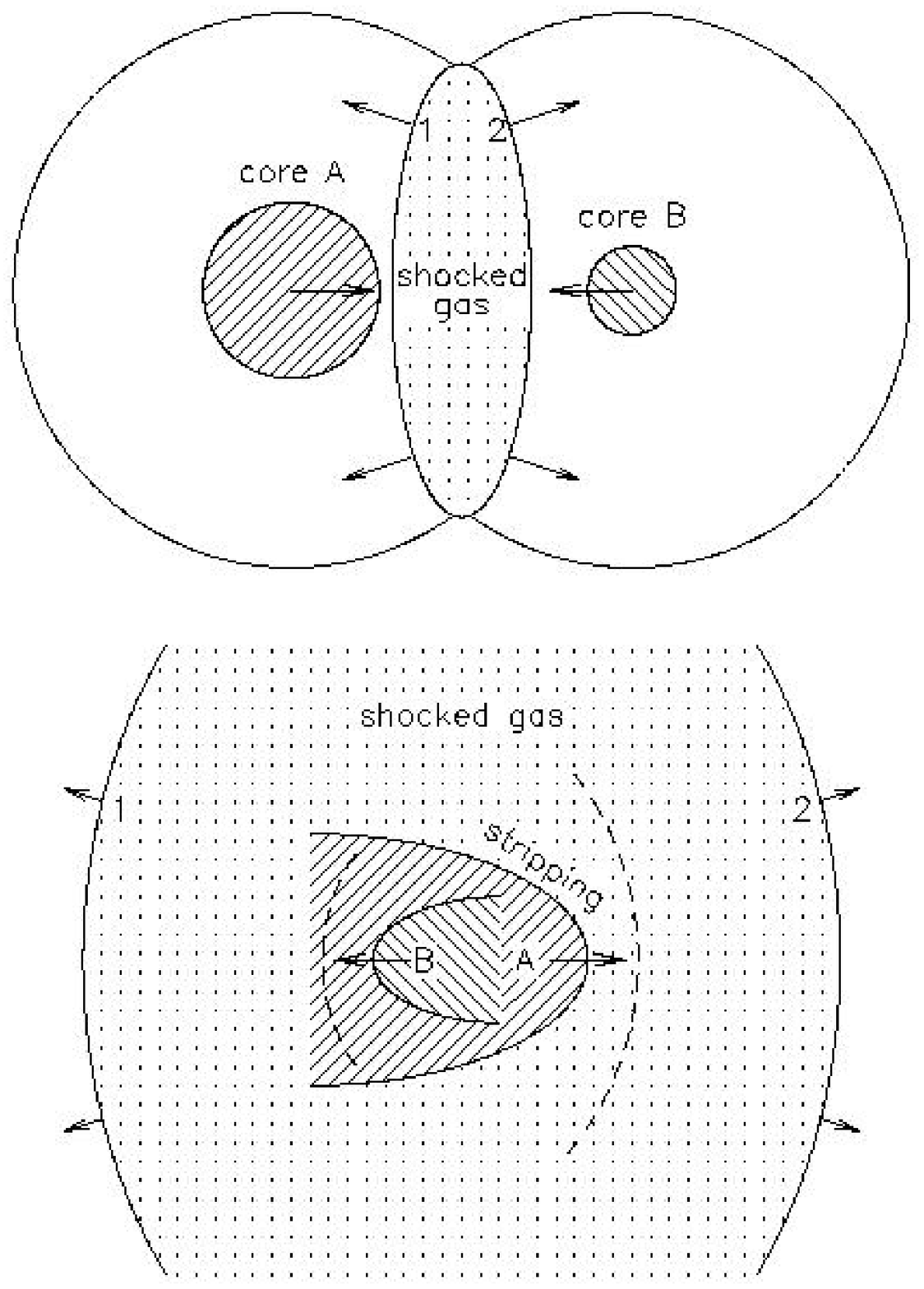}}
\caption{The 0.3--10 keV band ACIS image of A2142, binned to $2''$,
shows two sharp elliptical surface brightness edges northwest and south of the
cluster peak (see Markevitch et al. for details).
Right: One model for the origin of the features in A2142, as
suggested by Markevitch et al., is shown.
Shock fronts 1 and 2 have crossed the cluster and are now in the
cluster outskirts. The shocks are unable to penetrate the dense cores
that continue to move through the shocked gas. The cores may develop
additional leading shock fronts (dashed lines). The edges correspond
to the leading edges of the cold cores of the original merging
clusters.}
\label{a2142_image}
\label{a2142_model}
\end{figure}

The Chandra 0.3-10 keV band ACIS image of A2142, shown in
Fig.~\ref{a2142_image}, exhibits two sharp surface brightness edges --
one lies $\sim 3'$ northwest of the cluster center (seen earlier in
the ROSAT image) and a second lies $\sim 1'$ south of the
center.\cite{mark2000} Markevitch et al. derived the gas density, gas
temperature, and gas pressure distributions across the edges in the
cluster.\cite{mark2000} The gas temperature distribution shows a sharp
and significant {\em increase} as the surface brightness (gas density)
{\em decreases}. For the southern edge, the gas temperature rises by
about a factor of 2 from $\sim 5$ keV to $\sim 10$ keV. For the
northwestern edge, the surface brightness is lower and the
uncertainties are larger, but the temperature change is comparable.
The gas density changes across the edges compensate, within the
uncertainties, for the temperature increases so that the gas pressures
across the edges are equal.

One conclusion from the Chandra observation of A2142 in
unambiguous. The edges in A2142 do {\em not} arise from shocks. If
these edges were shocks, the gas temperature in front of the shock
(i.e. away from the cluster center) would be lower than that behind
the shock. This is exactly the opposite of what is
found.\cite{mark2000} The observed features are called ``cold
fronts''.\cite{vik2001a}

One suggestion for the origin of the A2142 structures is that they
arise from the merger of two systems and that the dense cores have
survived the merger. We observe A2142 as it would appear after the
shock fronts have passed by each of the dense cores (see
Fig.~\ref{a2142_model}). The outer, lower density gas has been shock
heated, but the dense cores remain ``cold''.\cite{mark2000} Each sharp
edge is then a boundary of a ram pressure-stripped subcluster
remnant.

\subsection{Cluster Physics and Edges}

A3667, a moderately distant cluster ($z=0.055$ ; 1.46 kpc per arcsec),
was a second candidate system with an edge, seen by ROSAT, that was
expected to exhibit a shock front.\cite{mark1999} However,
as with A2142, this feature also is the boundary of a dense cold
cloud, a merger remnant, as it traverses the ICM.\cite{vik2001a,vik2001b}

\begin{figure}
\centerline{\includegraphics[width=0.85\textwidth]{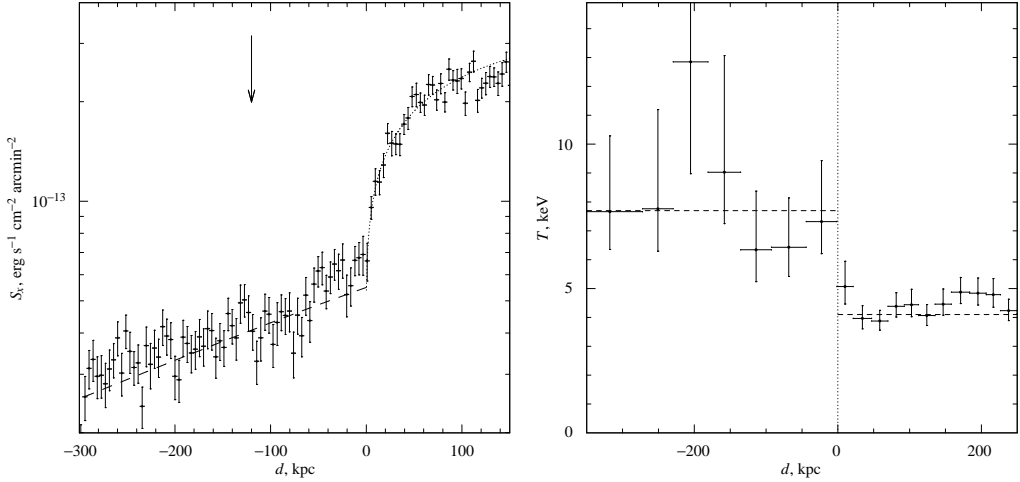}}
\caption{Left: The 0.5-2.0 keV surface brightness profile for A3667
extracted in elliptical regions across the cold front. The large,
sharp drop is clearly seen. The dashed line is the ROSAT PSPC fit to
the outer surface brightness distribution and agrees well with the
Chandra observation. The dotted curve is a fit to a spheroid with a
sharp boundary. As discussed by Vikhlinin et al. (2001a), the excess
at distances of 0-50 kpc in front of the edge represents gas that
accumulates in the stagnation region. Right: The temperature profile
across the cold front. The temperature {\em increases} from $\sim 4$
keV to $\sim 8$ keV across the front. }
\label{a3667_profs}
\end{figure}

As Vikhlinin et al.  showed, the edge is accurately modeled as a
spheroid (see left panel of Fig.~\ref{a3667_profs}).\cite{vik2001a}
From the surface brightness profile, converted to gas density, and
precise gas temperatures, the gas pressure on both sides of the cold
front can be accurately calculated . The difference between the two
pressures is a measure of the ram pressure of the ICM on the moving
cold front. Hence, the precise measurement of the gas parameters
yields the cloud velocity. Using the ratio of the pressures in the
free streaming region and the stagnation point (immediately in front
of the cold front), the factor of two difference in pressure across
the front yields a Mach number of the cloud of $1\pm0.2$ ($1430\pm290$ km
s$^{-1}$).\cite{vik2001a}

In addition to the edge, a genuine weak shock is detected, consistent
with the above calculated velocity.  The distance between the cold
front and the weak shock ($\sim350$ kpc) and the observed gas density
jump at the shock (a factor of 1.1-1.2) yield the shock's propagation
velocity, $\sim 1600$ km s$^{-1}$, consistent with that derived
independently for the cold front.\cite{vik2001a}

The A3667 observation provides important information on the
efficiency of transport processes in clusters. As the temperature and
surface brightness profiles show (see Fig.~\ref{a3667_profs}), the edge
is quite sharp. Quantitatively, Vikhlinin et al. found that the width
of the front was less than $3.5''$ (5 kpc). This sharp edge requires
that transport processes across the edge be suppressed,
presumably by magnetic fields. Without such suppression, the edge
should be broader since the relevant Coulomb mean free path for electrons
is about 13 kpc, several times the width of the cold front.\cite{vik2001a} 

Finally, Vikhlinin et al. observed that the cold front appears sharp
only over a sector of about $\pm 30\deg$ centered on the direction of
motion, while at larger angles, the sharp boundary
disappears.\cite{vik2001b} The disappearance can be explained by the
onset of Kelvin-Helmholtz instabilities, as the ambient gas flows past
the moving cold front. To explain the observed extent of the sharp
boundary, the instability must be partially suppressed, e.g., by a
magnetic field parallel to the boundary. Its required strength is
$7-16\mu$G. Such a parallel magnetic field may be drawn out by the
flow along the front. This measured value of the magnetic field in the
cold front implies that the pressure from magnetic fields is small
(only 10-20\% of that of the thermal pressure) and, hence, supports
the accuracy of cluster gravitating mass estimates derived from X-ray
measurements that assume that the X-ray emitting gas is in hydrostatic
equilibrium and supported by thermal pressure.\cite{vik2001b}

\section{The Realm of Galaxies}

Chandra observations of early type galaxies have brought new data and
some surprises. We first discuss M86, the prototype of a galaxy
undergoing ram pressure stripping of its hot gaseous corona as the
galaxy crosses the core of the Virgo cluster. Second we summarize the
Chandra observations of NGC507, the central galaxy in a group whose image
shows both the effects of plasma bubbles on its gaseous atmosphere and
sharp edges of unknown origin.

\subsection{M86=NGC4406}

\begin{figure}
\centerline{\includegraphics[width=0.5\textwidth]{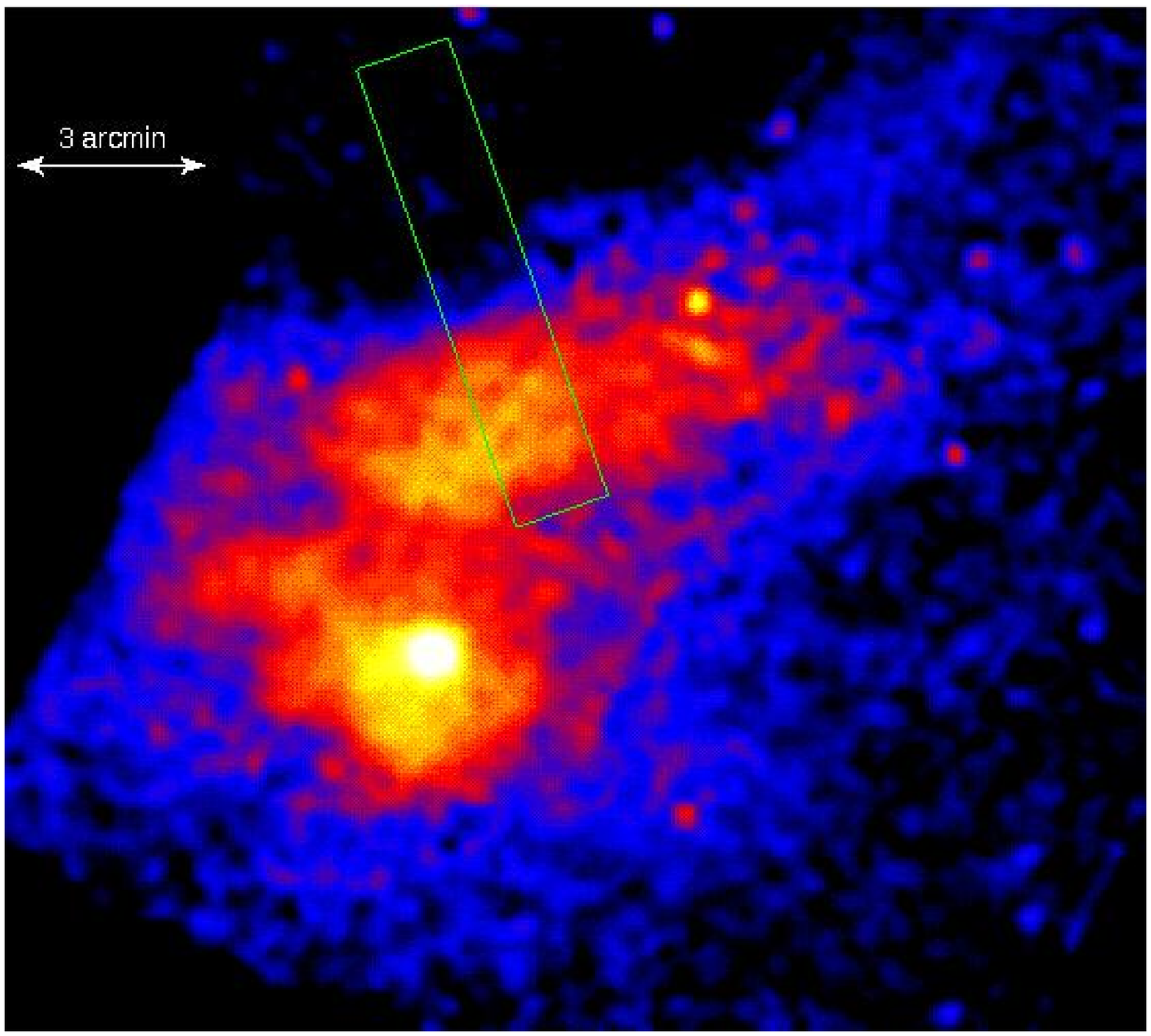}\includegraphics[width=0.5\textwidth]{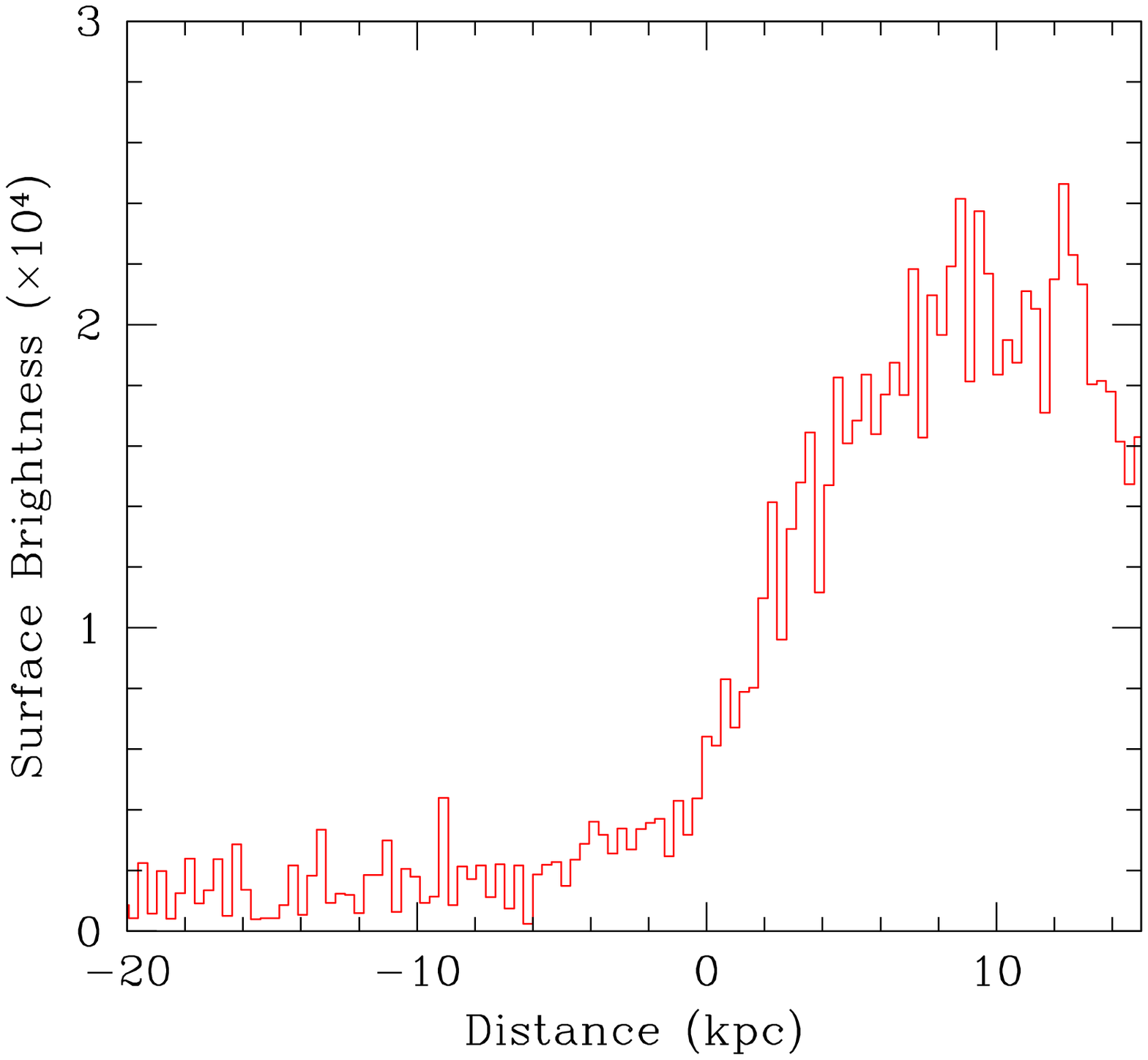}}
\caption{The 0.5-2.0 keV surface brightness image of M86. The
green rectangle indicates the region used to generate the projection
shown in the right portion of the figure. The edge to the north is
extremely sharp with the surface brightness and gas temperature
changing on a scale of less than $\sim 5$ kpc. }
\label{m86}
\end{figure}

X-ray emission from M86 was first observed with the Einstein
Observatory and subsequently by ROSAT. Its unusual ``plume'' was
explained as a ram pressure stripped galactic
corona.\cite{wrf1979,fabian1980,ros1,ros2} The Chandra observation
shows a remarkably sharp boundary between the ram pressure stripped
corona and the hotter Virgo cluster. As Fig.~\ref{m86} shows, the cool
($\sim0.6$ keV) corona is separated from the hot ($\sim3$ keV) Virgo
ICM by less than about 5 kpc, comparable to the width of the cold
front in A3667. The ``edge'' here differs from that in A3667 since M86
is traversing the Virgo ICM at supersonic velocity and the edge lies
to the rear or side of M86 as it moves towards us and to the
southeast.

\begin{figure}
\centerline{\includegraphics[width=0.75\textwidth]{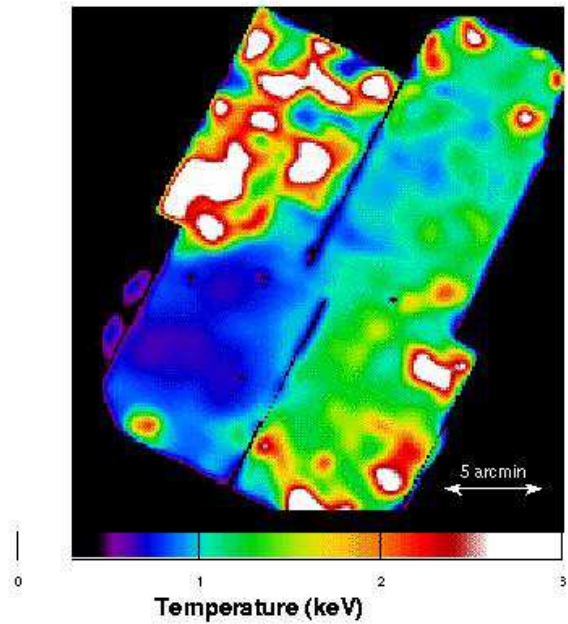}}
\caption{An updated temperature map of M86 combining data from
different chips and two different pointings. The agreement in
temperature across the different ACIS chips and between the two
observations is excellent. Point sources are excluded (small dark
spots in temperature map). The galaxy core and central region of the
plume are cool ($kT\sim0.7$ keV). The temperature boundary in the
northeast between the
plume and the Virgo cluster gas is sharp with the temperature varying
from $\sim0.7$ keV to $\sim2-3$ keV.}
\label{m86_tmap}
\end{figure}

The precise origin for the structure of M86 remains unclear. As was
pointed out by Toniazzo \& Schindler none of their simulations viewed
from any orientation exactly matched the morphology of M86.\cite{sim}
Perhaps there is physics that must be added to the simulations and, as
in clusters, magnetic fields play an important role in shaping the gas
and temperature distribution.

\subsection{NGC507 - the central galaxy in a group}

\begin{figure}
\centerline{\includegraphics[width=0.5\textwidth,bb=130 220 450 550,clip]{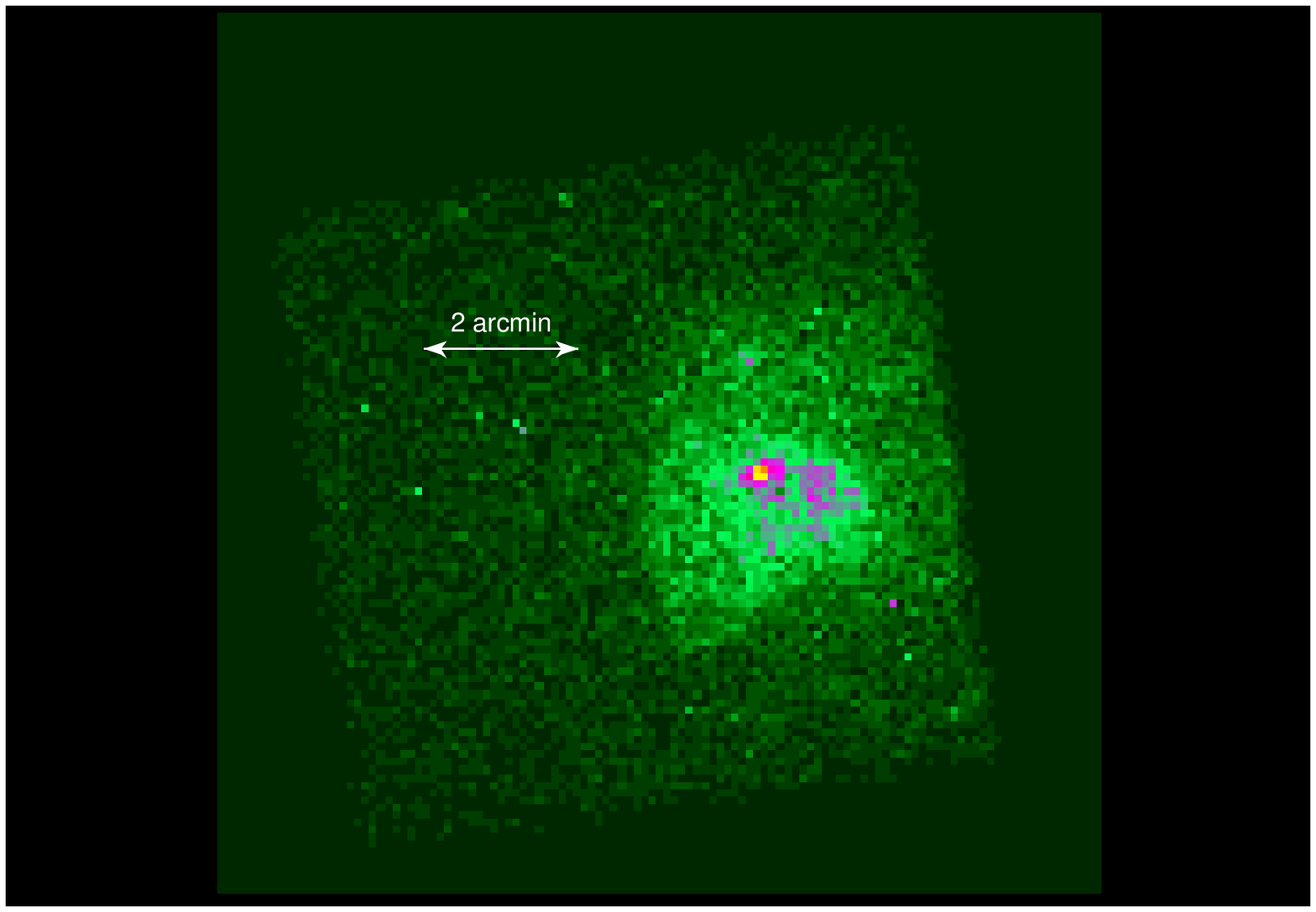}
\includegraphics[width=0.5\textwidth,bb=23 23 573 556,clip]{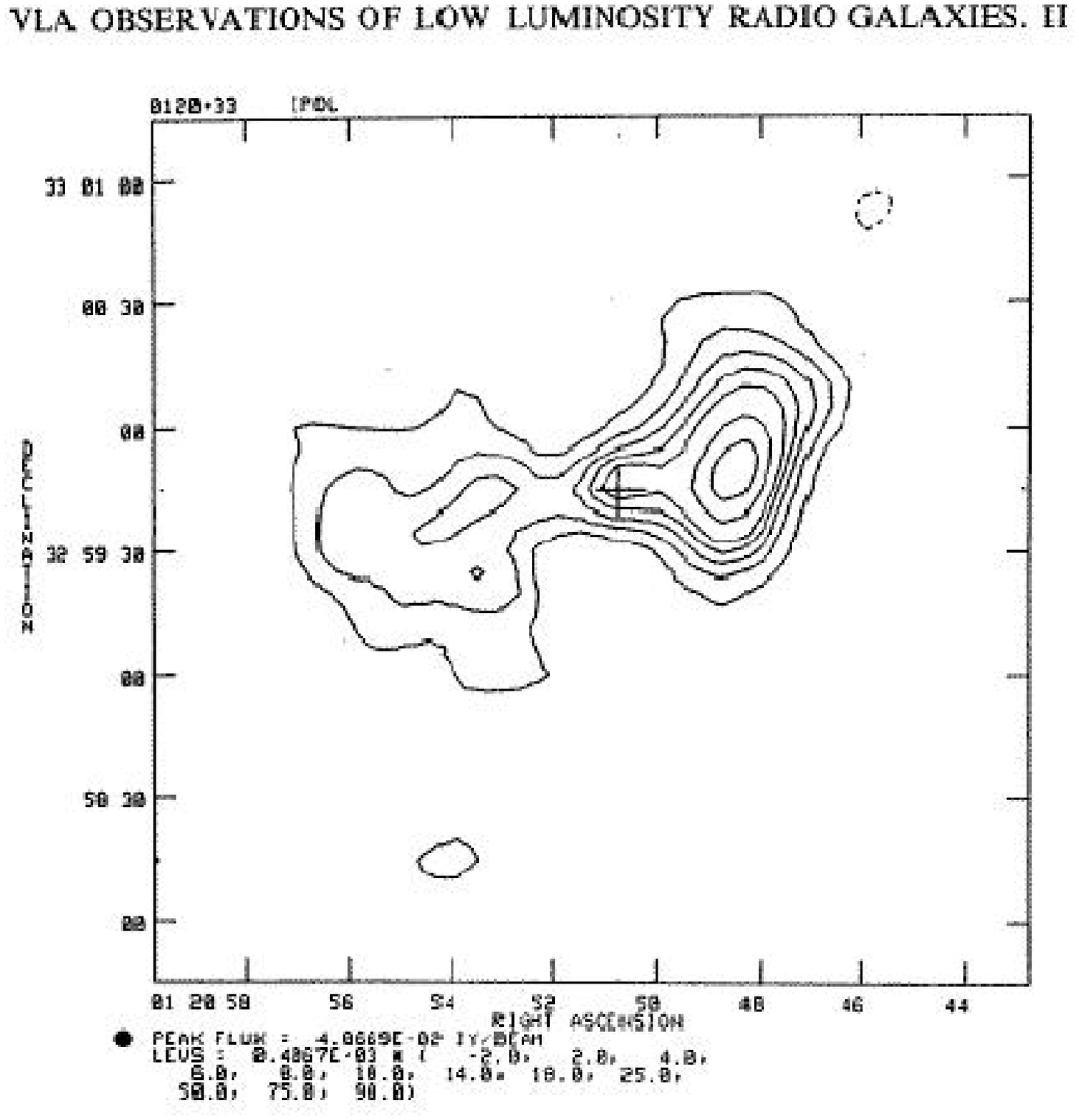}}

\vspace*{0.2in}
\centerline{\includegraphics[width=0.5\textwidth]{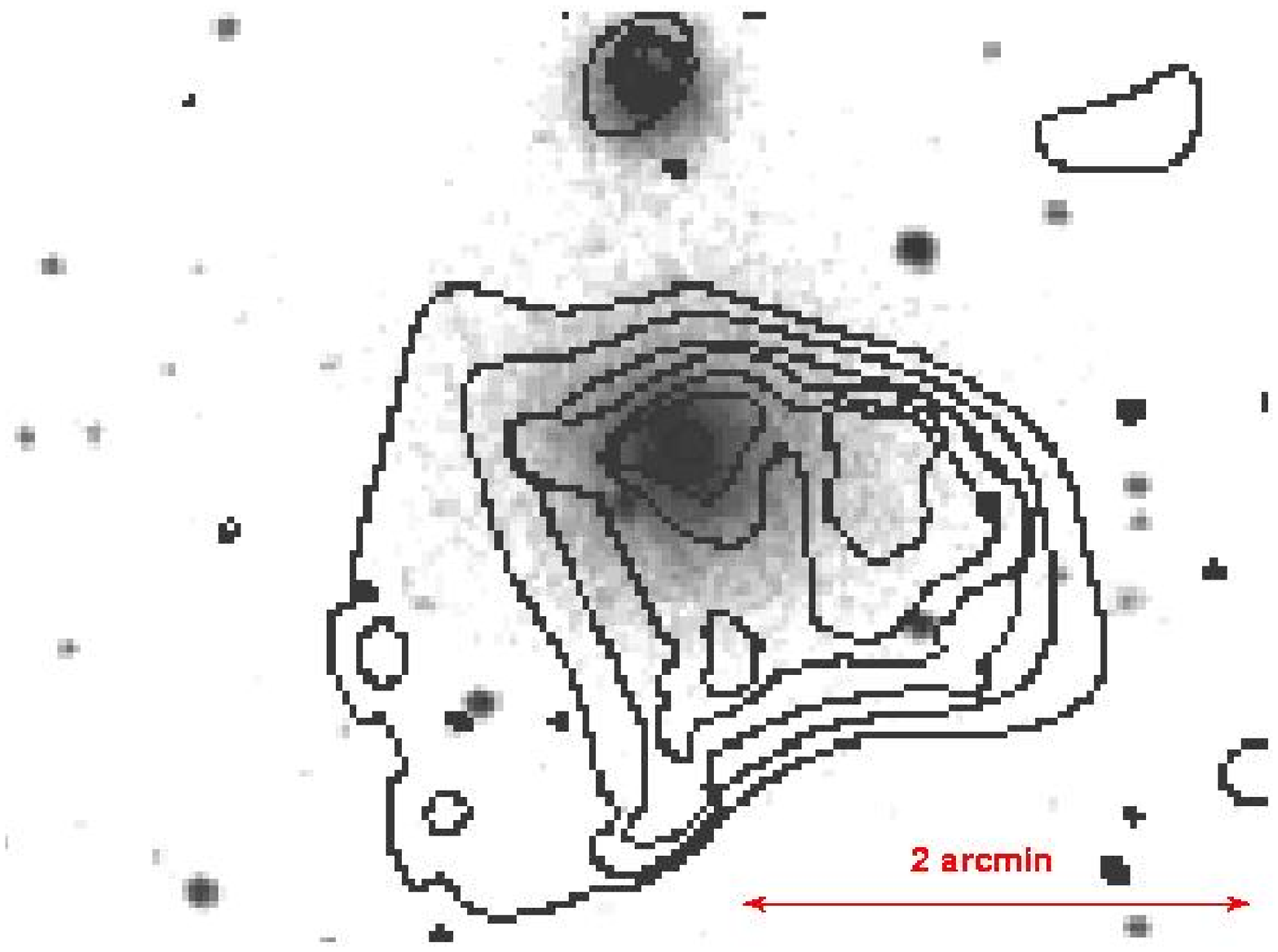}\includegraphics[width=0.5\textwidth,bb=20
165 555 640,clip]{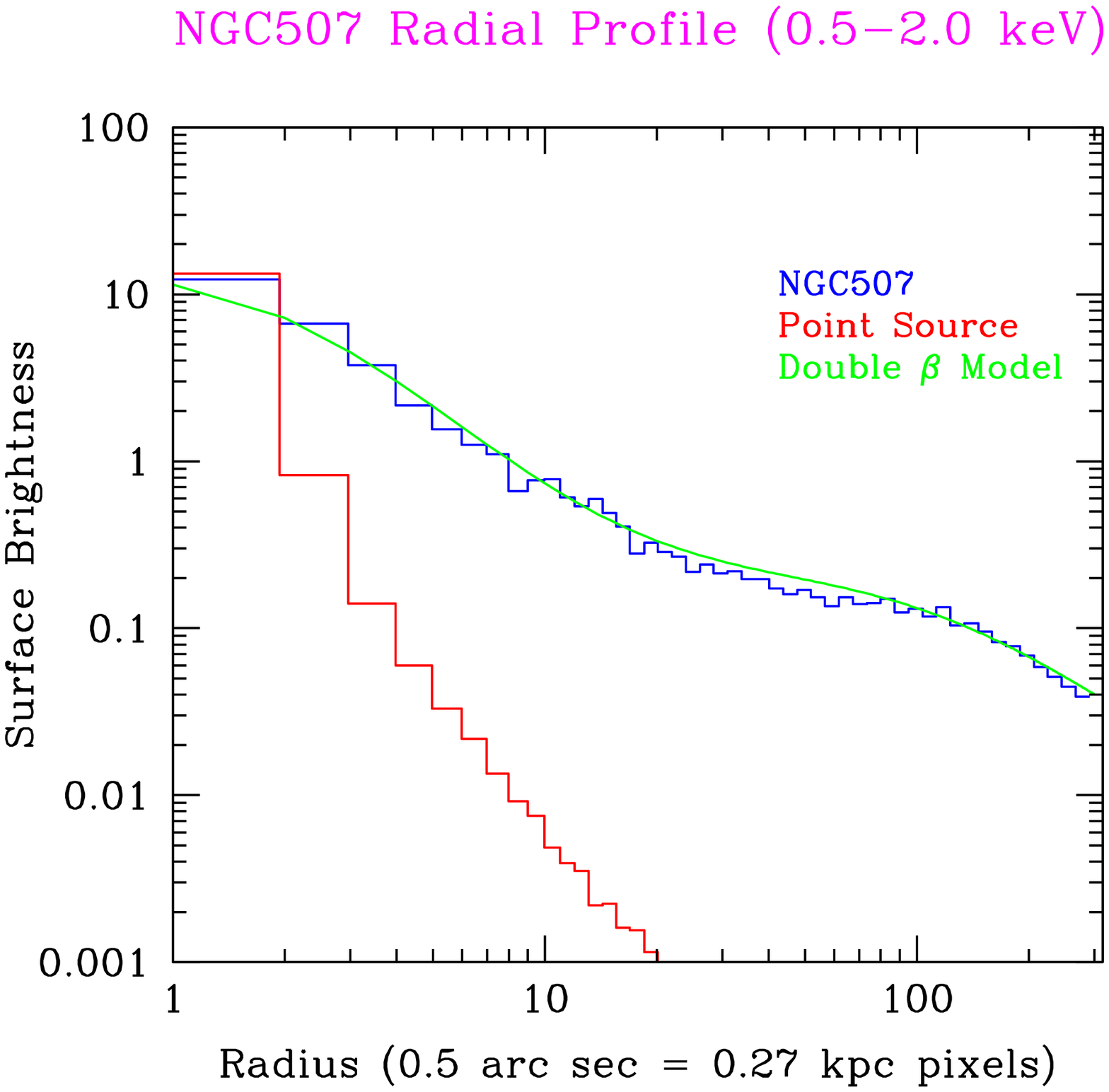}}

\caption{Top Left: The 0.5-2.0 keV surface brightness distribution of
NGC507. Top Right: The VLA radio map showing the central point source,
a jet emanating to the west and two radio lobes (see de Ruiter et
al. 1986).  The depression in the X-ray surface brightness to the west
of the galaxy peak coincides with the western radio lobe. Bottom Left:
The X-ray contours overlayed on the optical image show the two X-ray
``peaks''.  The eastern peak is centered on the optical galaxy while
the western peak has no optical counterpart, but is probably produced
by the intervening western radio lobe which produces the depression
between the two peaks by displacing X-ray emitting gas. Bottom Right:
Radial profile (0.5-2.0 keV) of the emission around NGC507 (blue
histogram) along with a point source (red histogram) and a double
$\beta$ model (green). Any contribution by an active nucleus to the
total emission from the central region must be small.  }
\label{n507}
\end{figure}

NGC507 is the central galaxy in a rather nearby ($z=0.016458$),
group that has been studied extensively in
X-rays.\cite{kim1995,matsumoto1997,buote1998,fukazawa1998} The galaxy
is the site of a weak B2 radio source (luminosity of about $10^{37}$
ergs s$^{-1}$).\cite{ruiter} The Chandra X-ray image, shown in
Fig.~\ref{n507} (top left), covers only the central, high surface
brightness region of the group around NGC507.  The 0.5-2.0 keV surface
brightness distribution shows sharp edges to the southwest, southeast
and north, reminiscent of those in the clusters A2142 and A3667. In
addition to the edges, there are two X-ray peaks. The first, to the
east, coincides with the nucleus of NGC507 (see X-ray contours
overlayed on the optical image in the lower left panel of
Fig.~\ref{n507}). A second peak, $1'$ to the west has no optical
counterpart. However, comparing the X-ray and the radio map (upper right
panel of Fig.~\ref{n507}) shows that the western
radio lobe lies precisely in the surface brightness trough between the
nucleus and the peak to the west. Thus, it seems  likely that the
radio lobe, possibly a buoyant bubble, has displaced X-ray emitting
gas generating a trough in the X-ray surface brightness distribution.

The origin of the peculiar sharp surface brightness discontinuities
around NGC507 is unclear. The bright emission is well fit by a thermal
model with gas temperatures near 1 keV, consistent with the mean ASCA
temperature of $1.10\pm0.05$ keV.\cite{matsumoto1997} The emission
from the central region is resolved and hence the contribution from a
central AGN is relatively small (see Fig.~\ref{n507} bottom right
panel).  Perhaps the X-ray surface brightness features arise either
from motion of NGC507 and its dark halo within the larger group
potential as suggested for the multiple edges in
clusters.\cite{mark2001} Alternatively, perhaps the edges are relics
of the merger process in a group as dynamical friction causes the
merging of smaller galaxies with the central dominant NGC507. Thus we
could be observing tracers of the process that forms fossil groups
and OLEGS, systems near the end points of their dynamical evolution
that contain a single optically bright
galaxy.\cite{ponman,vik1999}

\section{Conclusions}

We did not anticipate surprises from Chandra's high angular resolution
observations of clusters and early type galaxies.  Instead of
confirming our prejudices, Chandra has brought us a wealth of new
information on the interaction of radio sources with the hot gas in
both galaxy and cluster atmospheres. We see ``edges'' in many systems
with hot and cold gas in close proximity and have been able to extract
important new parameters of the ICM from their study.  We have only
barely begun to digest the import of the Chandra cluster and galaxy
observations.  We can only expect the unexpected as Chandra
observations continue and as our understanding of how best to use this
new observatory matures.

We acknowledge support from NASA contract NAS8 39073, NASA grants
NAG5-3065 and NAG5-6749 and the Smithsonian Institution.

\end{document}